\documentstyle[12pt,aaspp4]{article}
\def\delchi{{$\Delta\chi^{2}$}}
\def\etal{{et al.}}

\def\asca{{\it ASCA}}

\def\chisq{{$\chi^{2}$}}

\def\Msun{\hbox{$\rm\thinspace M_{\odot}$}}

\begin{document}

\title {THE PROPERTIES OF THE RELATIVISTIC IRON K-LINE IN NGC 3516}

\author {K. Nandra \altaffilmark{1, 2},
	I.M. George \altaffilmark{1,2},
	R.F. Mushotzky \altaffilmark{1}, 
	T.J. Turner \altaffilmark{1,3}, 
	T. Yaqoob \altaffilmark{1,2}}

\altaffiltext{1}{Laboratory for High Energy Astrophysics, Codes 660.2 \& 662, 
	NASA/Goddard Space Flight Center,
  	Greenbelt, MD 20771}
\altaffiltext{2}{Universities Space Research Association}
\altaffiltext{3}{University of Maryland at Baltimore County}

\slugcomment{Submitted to 
{\em The Astrophysical Journal Letters}}

\begin{abstract}

We present an analysis of the relativistic iron K$\alpha$ line in the
Seyfert 1 galaxy NGC 3516, based on a continuous, five-day ASCA
observation.  The broad profile which has been found in several other
AGN is confirmed in NGC 3516 with unprecedented signal-to-noise
ratio.  Disk-line models with either a Kerr or Schwarzschild metric
fit the integrated profile, but both require emission very strongly
concentrated in the inner disk.  We find tentative evidence for the
line signatures of Ni K$\alpha$ and/or Fe K$\beta$.  The continuum
flux varied by $\sim 50$~per cent during the observation and
time-resolved analysis shows that the line also changes. The line core
seems to follow the continuum, but the blue wing is unrelated and
shows a greater amplitude (factor $\sim 2$) of variability. The red
wing is formally consistent with a constant but appears to be
correlated with the blue wing.  We interpret this as evidence for
independent variability of the broadest parts of the line.  There also
appears to be an absorption feature in the profile, consistent with
resonance scattering in infalling material. This variable feature may
be the signature of material being accreted by the central black hole.

\end{abstract}


\section{INTRODUCTION}
\label{Sec:Introduction}

A long \asca\ exposure of the Seyfert 1 galaxy MCG-6-30-15 revealed a
broad, redshifted profile to the iron K$\alpha$ line (Tanaka \etal\
1995). The line is thought to arise from the inner regions of an
accretion disk close to the central black hole, the profile being the
result of extreme Doppler and gravitational shifts (e.g., Fabian
\etal\ 1995). ``Snapshot'' observations of other Seyferts have
shown these broad lines to be common and defined the mean properties
(Nandra \etal\ 1997b, hereafter N97b; Reynolds 1997).  Furthermore,
Iwasawa \etal\ (1996, hereafter I96) have presented evidence that the
profile of MCG-6-30-15 is variable, being extremely broad and
redshifted in the ``deep minimum'' state (see also Reynolds \&
Begelman 1997; Young, Ross \& Fabian 1998; Iwasawa et al. 1999).

It is of great interest that the line profile is observed to vary in
this way, and indeed at all. The lines typically arise around $20
R_{\rm g}$ (Tanaka et al. 1995; N97b) and should therefore be highly
responsive to continuum variations: the expected lag due to
reverberation is of order $\sim 1000$s.  In the time taken to
accumulate a typical \asca\ spectrum ($\sim 20-40$~ks) we would
therefore expect the line profile to relax to its mean state and
appear similar at all times.  I96 suggested that, rather than due to
reverberation, the profile changes might be due to a changing pattern
of illumination of the disk (see also Weaver \& Yaqoob 1998; McKernan
\& Yaqoob 1999; Reynolds et al. 1999).  Here we report on a long
observation of the Seyfert 1 galaxy NGC 3516 (z=0.009), which is known
to be variable and have a broad, redshifted iron K$\alpha$ line (Kriss
\etal\ 1996; N97 a,b,c), offering the opportunity for performing
analysis similar to that of MCG-6-30-15.

\section{OBSERVATIONS}

The \asca\ observation began on 1998-Apr-12 at 21:44:50 and lasted for
$\sim 4.5$ days.  We used data from both Solid--state Imaging
Spectrometers (SIS0/SIS1), and the two Gas Imaging Spectrometers
(GIS2/GIS3). The SIS data were collected in 1-CCD readout mode using
{\tt FAINT} telemetry mode. The data were analyzed according to the
methods described in Nandra \etal\ (1997a,b) and references therein.
The total exposure times after screening were 152 ks for the SIS
detectors and 186 ks for the GIS, making this one of the longer \asca\
exposures of a single object.  The mean count rates for the
observation were $0.978 \pm 0.003$, $0.806 \pm 0.002$, $0.587 \pm
0.002$ and $0.724 \pm 0.002$ for SIS0, SIS1, GIS2 and GIS3 (SIS:
0.5-10 keV; GIS: 1-10 keV). The average 2-10 keV flux, determined
using a power-law model, was $4.7 \times 10^{-11}$~erg cm$^{-2}$
s$^{-1}$, which lies between the flux states described by N97c.

\section{MEAN LINE PROPERTIES}

\begin{figure}

\centering 
\epsfxsize=0.95\textwidth
\epsffile{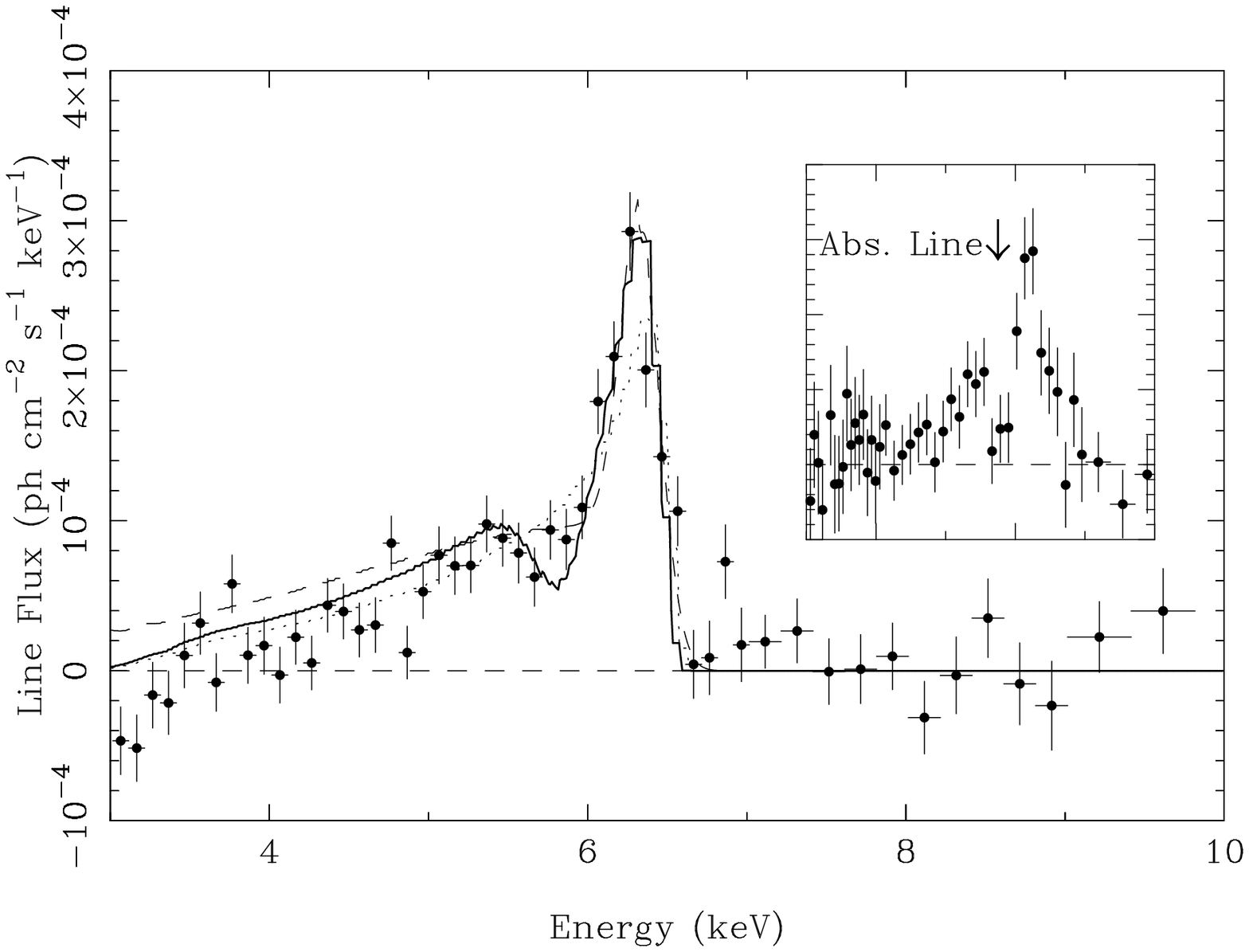}

\caption{
SIS line profile for the whole observation, created by interpolating a
local power-law continuum into the 4-7 keV region. The solid line
shows our best fit model for this profile, that of an accretion disk
around a rotating (Kerr) black hole, with an absorption line -
presumably also due to iron but redshifted - in the low-energy wing of
the line.  This provides a better fir to the data than a line from the
disk line alone (dotted line), even when an additional narrow
component is allowed (dashed line).  Inset shows the line profile from
the parts of the observation (designated ``P4'' and ``P5'': see text
and Fig~\ref{fig:lc}) where this absorption feature was
found to be the strongest.  The arrow marks its location.
\label{fig:profile}
}
\end{figure}

We first analyzed the integrated spectrum.  The X-ray spectrum of NGC
3516 is absorbed by a column of partially ionized gas (e.g., Kriss
\etal\ 1996) and we restricted our initial analysis to the 3-10 keV
band to minimize its effects. We found no evidence for an iron K-edge
from the ionized gas.  Fig.~\ref{fig:profile} shows the line profile
of NGC 3516 derived from the SIS data only, derived by adopting a
local power law fit in the 3-4 and 7-10 keV range to the SIS+GIS data.
It is clearly broad and red-shifted, and shows a very similar shape to
MCG-6-30-15 (Tanaka \etal\ 1995), the mean for Seyfert 1 galaxies
(N97b) and previous observations of NGC 3516 (N97c).  The profile can
be modeled adequately with two gaussians, a narrow ``core'' and broad,
redshifted ``wing'', with parameters detailed in
Table~\ref{tab:line}. The fit gave a rather flat photon index compared
with previous \asca\ observations (N97c), with $\Gamma=1.56\pm 0.04$
and an acceptable \chisq\ of 1703.0/1736 d.o.f.

\begin{deluxetable}{lll}

\tablecaption{Line parameters derived from fitting the mean spectrum.
\label{tab:line}}

\tablehead{
\colhead{Double gaussian:} &
\colhead{Core} &
\colhead{Wing}
}
\startdata
$\Gamma$                & $1.56 \pm 0.04$ & \nodata \nl
$E_{K\alpha}$  		& $6.31^{+0.04}_{-0.04}$ & $5.36^{+0.24}_{-0.40}$ \nl
$\sigma_{K\alpha}$  	& $0.09^{+0.07}_{-0.09}$ & $1.16^{+0.45}_{-0.29}$ \nl
$I_{K\alpha}$       	& $0.74^{+0.18}_{-0.16}$ & $2.78^{+1.82}_{-0.95}$ \nl
$W_{K\alpha}$       	& $140^{+34}_{-30}$  & $470^{+310}_{-160}$  \nl
$\chi^{2}$/dof  	& 1703.0/1736            & \nodata \nl
 & & \nl \hline
Disk line: & Schwarzschild & Kerr \nl \hline
$\Gamma$       	& $1.53^{+0.03}_{-0.04}$ & $1.52^{+0.05}_{-0.02}$  \nl
$E_{dl}$       	& $6.40^{+0.07}_{-0.0P}$ & $6.54^{+0.05}_{-0.09}$ \nl
$q$ 		& $8.0^{+2.0P}_{-3.3}$	 & $2.71^{+0.12}_{-0.15}$ \nl
$R_{\rm i}$ 	& $6.0$ (F)              & $2.8^{+1.0}_{-1.6P}$ \nl
$R_{\rm o}$ 	& $80^{+\infty P}_{-71}$  & $400$ (F) \nl
$i$ 		& $35^{+1}_{-2}$         & $0^{+19}_{-0P}$ \nl
$I_{dl}$       	& $2.8^{+0.2}_{-0.2}$    & $2.4^{+0.4}_{-0.4}$ \nl
$W_{dl}$       	& $620^{+40}_{-40}$      & $530^{+80}_{-80}$ \nl
$\chi^{2}$/dof 	& 1696.7/1736            & 1703.4/1736 \nl
\enddata

\tablecomments{
$\Gamma$ is the continuum photon index;
$E_{K\alpha}$ is the centroid energy of the gaussian in keV;
$\sigma_{K\alpha}$ is the width (keV); $I_{K\alpha}$ is the 
intensity of the gaussian in units of $10^{-4}$ ph cm$^{-2}$ s$^{-1}$; 
$W_{K\alpha}$ is the equivalent width in eV;
$E_{dl}$ is the rest energy of the disk line (keV);
$q$ is the emissivity index (see text);
$R_{\rm i}$ is the disk inner radius; 
$R_{\rm o}$ is the disk outer radius;
$i$ is the disk inclination;
$I_{dl}$ is the intensity of the disk line ($10^{-4}$ ph cm$^{-2}$ s$^{-1}$);
$W_{dl}$ is the equivalent width of the disk line (eV);
(F) denotes that the parameter has been fixed;
P denotes that the parameter ``pegged'' at a limiting value
}
\end{deluxetable}

We next fitted the data with the Schwarzschild disk line model of
Fabian \etal\ (1989) .  Such disk line models are rather complex, with
considerable degeneracy which makes the \chisq-fitting process
susceptible to traps in local minima.  The solutions we show in
Table~\ref{tab:line} are the best we have found after searching
parameter space extensively. Determination of accurate error bars in
these circumstances is severely complicated by the model degeneracies,
however, and these should be treated with caution.  Fits to the
Schwarzschild model provided an excellent fit to the data
(\chisq=1696.7/1736 d.o.f) - improving on the double-gaussian model by
\delchi=6.3 for the same number of free parameters. Both the energy of
the line and the inner radius pegged at their minimum values (6.34 keV
and 6.0 $R_{\rm g} = R_{\rm ms}$) in this fit. To improve the
stability of the fitting, the inner radius was therefore fixed at
$R_{\rm ms}$ when the parameters in Table~\ref{tab:line} were
determined. This low inner radius and the extraordinarily steep
emissivity index of $q=8.0$ indicate that the line emission is very
strongly concentrated in the central regions, where the most extreme
relativistic effects operate.  Another way of enhancing these
relativistic effects is to allow the disk to extend closer to the
black hole, which is possible when the hole is rotating. Fits to the
Kerr model of Laor (1991) are also given in Table~\ref{tab:line}.  In
this case the outer radius cannot be constrained and was fixed at the
maximum value allowed by the implementation of the model ($400 R_{\rm
g}$). The \chisq of 1703.4/1736 dof for is slightly worse than the
Schwarzschild case but the parameters, in particular the emissivity
law, are more in line with our prejudices about the source
geometry. For a disk centrally-illuminated by a point source we expect
q$\sim 0-3$, for example.  The Kerr model is plotted as the dotted
line in Fig.~\ref{fig:profile}.

Weaker lines from iron K$\beta$ and Ni K$\alpha$ are also expected
from the accretion disk (e.g., George \& Fabian 1991), and indeed
Fig.~\ref{fig:profile} shows some evidence for excess flux around the
expected energies of 7-7.5 keV.  We have added additional disk lines
to the Kerr and Schwarzschild models to represent emission from Fe
K$\beta$ and Ni K$\alpha$. The energies were fixed at the neutral
values of 7.06~keV and 7.47~keV (rest frame) respectively, the fluxes
were left free, but all other disk line parameters were tied to those
of Fe K$\alpha$. For the Kerr model, we find a decrease in
fit-statistic of only \delchi=2.2 for Fe K$\beta$, but a significant
improvement (\delchi=5.5) for Ni K$\alpha$. The EW of these lines are
unconstrained, but the best-fit values are 60 eV and 80 eV
respectively (c.f. 560 eV for Fe K$\alpha$). For the Schwarzschild
model both lines are highly significant, with K$\beta$ having
\delchi=6.3 and Ni K$\alpha$ \delchi=14.1, and EW=100 eV and 150 eV.
Using an optimistic error prescription of \delchi=2.3 we constrain the
ratio of Fe K$\beta$/Fe K$\alpha$= $0.14\pm 0.09$, consistent with the
theoretical expectation for neutral iron of 0.11 (Kikoin 1976).  The
Ni K$\alpha$/Fe K$\alpha$ ratio is $0.19\pm 0.07$, a little higher
than than the $\sim 0.07$ expected based on the fluorescence yields,
and solar abundances.

Although the disk line clearly accounts for the bulk of the flux and
provides a good fit, it is possible that there is a contribution from
more distant material, such as the BLR or molecular torus
(e.g. Ghisselini, Haardt \& Matt 1993; Krolik, Madau \& Zycki 1993).
Indeed adding a narrow line with fixed energy at 6.4 keV does improve
the fit in both cases with \chisq=1684.0/1735 dof for the
Schwarzschild model (\delchi=12.7; equivalent width, EW=
$50^{+30}_{-30}$ eV) and \chisq=1682.8/1735 dof for the Kerr
(\delchi=20.6; EW=$90^{+60}_{-50}$). Aside from a reduction in the
flux (and EW) the disk line parameters remain consistent within the
errors.  This model is shown as the dashed line in
Fig.~\ref{fig:profile}.  Another interesting possibility is that there
is an absorption component present in the red part of the line.
Adding such an absorption line to the Schwarzschild model gives
\chisq=1690.2/1734 dof (\delchi=6.5). For the Kerr model we obtain
\chisq=1675.1/1734 dof (\delchi=28.3) for a narrow absorption line and
a further improvement to \chisq=1670.4/1733 when the absorption line
is slightly broadened ($\sigma$=0.2 keV). This last model is the best
of all attempted against the integrated profile and is shown as the
bold line in Fig.~\ref{fig:profile}.  Further evidence for the
absorption feature is afforded by the time-resolved profiles, which we
explore next.

\section{LINE VARIABILITY}

\begin{figure}

\centering 
\epsfxsize=0.8\textwidth
\epsfysize=0.3\textheight
\epsffile{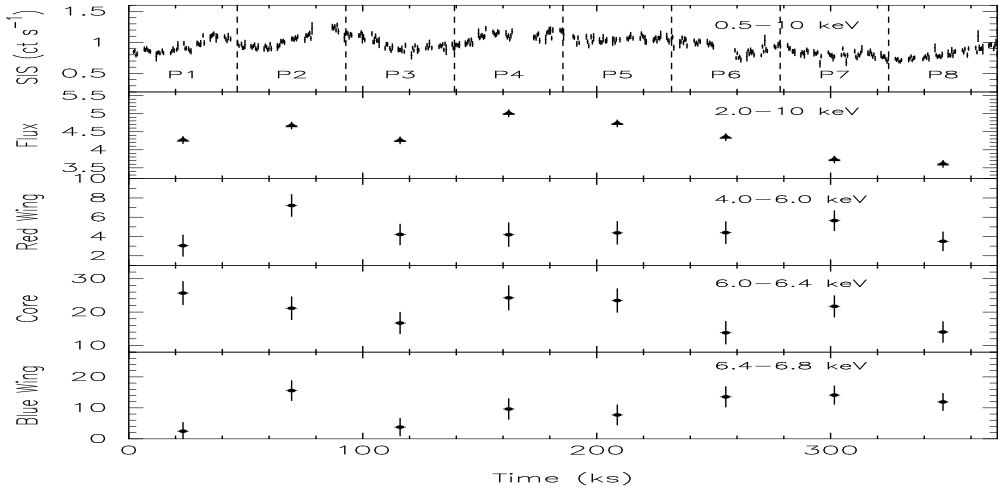}
\caption{(Top panel) SIS0/1 mean light curve in the
0.5-10 keV band, with a bin size of 512s.  Variations of up to $\sim
50$~per cent are observed. The vertical lines mark the periods which
were used for time-resolved spectroscopy (P1-P8). The remaining panels
show the results derived from these spectra. In descending order they
are the 2-10 keV continuum, and the excess flux above the continuum in
three line bands.  For the line fluxes, the \chisq\ values against a
constant hypothesis are 9.6, 13.1 and 18.4 respectively, for 7 degrees
of freedom in each case.  Neither the core nor the wing flux is
therefore consistent with a constant and though the red wing is
formally consistent with no variability, it appears strongly
correlated (at 95\% confidence) with the blue wing.
\label{fig:lc}}
\end{figure}

\begin{figure}

\centering 
\epsfxsize=0.95\textwidth
\epsffile{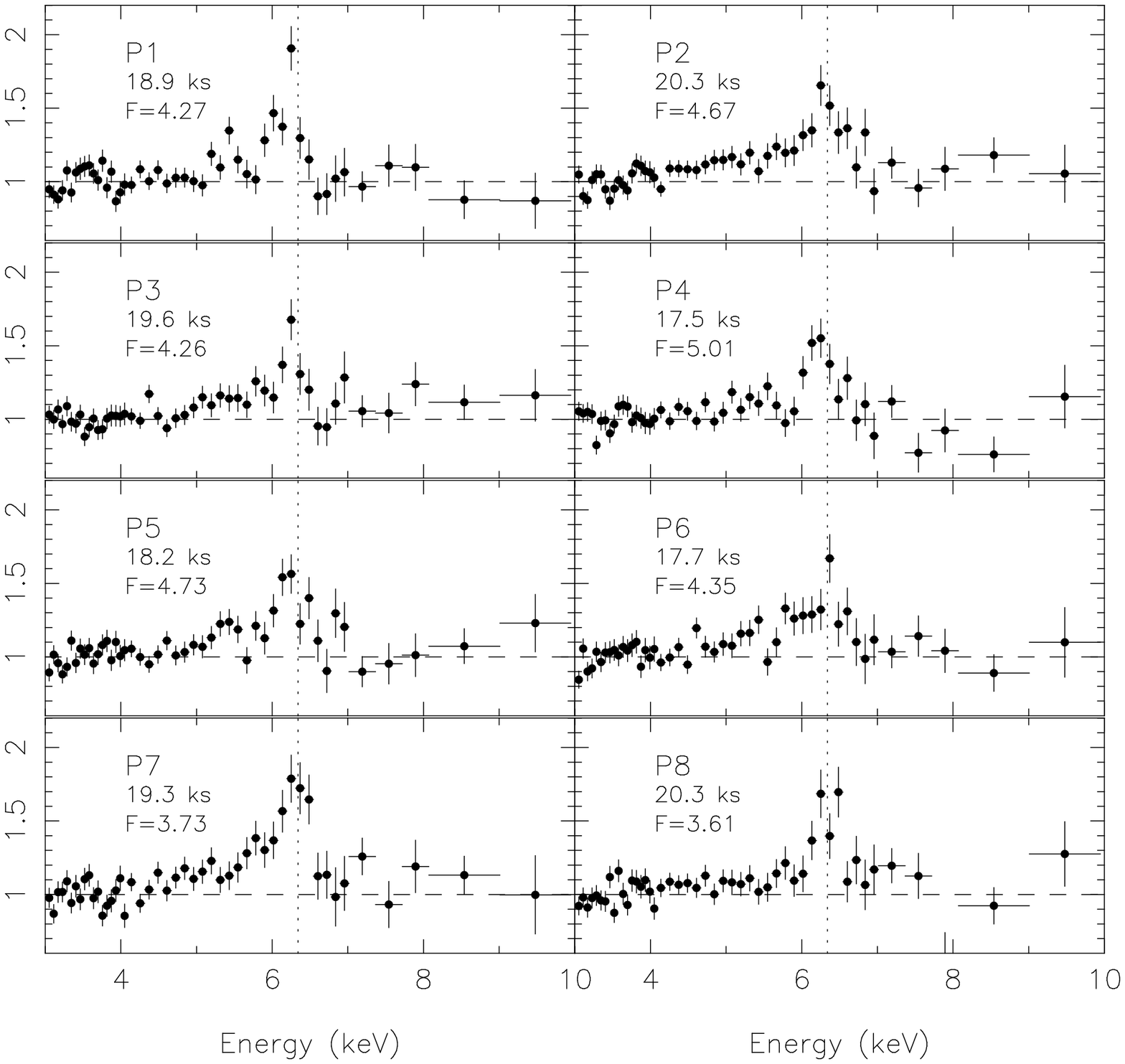}
\caption{
SIS data/model ratios for the individual segments marked in
Fig.~\ref{fig:lc}, derived from a power law fit the SIS+GIS spectra in
the 3-10 keV band. The 4-7 keV data were ignored in this fit then the
SIS data returned to derive the profile.  There are apparent changes
in the profile, but these are not easy to categorize from an
examination of this plot.
\label{fig:var_prof} }

\end{figure}

The continuum of NGC 3516 varied by a factor of $\sim 50$~per cent
during the observation (Fig.~\ref{fig:lc}), and we have searched for
variations in the emission line too.  We split the dataset into 8
segments, designated P1-P8, with equal durations (46.4 ks).  These are
marked on the light curve in Fig.~\ref{fig:lc}.  These segments have a
typical exposure of $\sim 20$~ks, which is just sufficient for
meaningful determination of the line parameters.  We used the same,
standard binning for all spectra.  The bins in each spectrum satisfy
our usual criterion of $>20$ ct/bin, but the standard binning method
ensures that when comparing line profiles, no spurious differences
appear due to different binnings.  We employed several different
methods of modeling the continuum to deconvolve the line: a) a power
law fit in the 3-4 and 7-10 keV bands (as above), b) a two-gaussian
fit to the 3-10 keV data, with energies and widths fixed at those for
the mean profile then the gaussians removed, and c) a photoionization
fit to the full band data, excluding the 4-7 keV band.  In practice
these all gave similar results.  We show the data/model ratios derived
from a) in Fig.~\ref{fig:var_prof}.  The peak seems to occur close 6.4
keV (shown by the vertical, dotted line) in most cases.  Although
individual profiles are noisy there does appear to be some variation
in the shape of the line.

We have quantified the profile variations by dividing the line into
three parts resolved in energy: the ``red wing'' (4-6 keV), ``core''
(6.0-6.4 keV) and ``blue wing'' (6.4-6.8 keV). We took the data model
ratios in these bins and converted them into line fluxes using the
fitted continuum. The resulting light curves are shown in
Fig.~\ref{fig:lc}. \chisq-tests against a constant show variability at
$\sim$ 80, 95 and 99 per cent confidences for the red wing, core and
blue wing.  The red and blue wing fluxes appear uncorrelated with the
continuum variations. A strictly proportional relationship between the
blue wing and continuum is ruled out at $\sim 99$~per cent confidence,
based on a \chisq\ test. Interestingly, however, the wings appear to be
correlated {\it with each other}. A linear (Pearson) correlation shows
a coefficient of $r=0.69$, significant at $\sim 95$ per cent
confidence.  The core is evidently variable with a poor \chisq of
$13.1$ for a constant intensity.  Assuming a 1:1 linear relationship
with the continuum gives a much better, and acceptable \chisq of
$10.0/8$ d.o.f. despite having 1 {\it fewer} degree of freedom.  This
offers some evidence that the continuum and line core are correlated
and vary in strict proportion, although the Pearson coefficient of
$r=0.49$ is not significant in this case.  We also fitted disk line
models to these individual spectra, but were not able to identify a
single parameter which accounts for the changes in profile.

Notwithstanding the complexity in determining the emission properties,
the absorption feature at $5.9$~keV found in the mean spectrum does
appear to be variable. We have fitted the SIS spectra only, which are
much more sensitive to the feature, with the ``template'' Kerr model
from the mean spectrum. The continuum spectral index and line flux
were allowed to vary, but all other line parameters were fixed at the
mean. We then added an absorption line at 5.89 keV, derived from the
mean spectrum, to the model. We found significant features (at $>99.5$
and $>95$ per cent confidence) in the P4 and P5 spectra (\delchi=9.6
and \delchi=4.7: see inset to Fig.~\ref{fig:profile}), but
insignificant improvements (\delchi$<2.2$) in all other
spectra. Examining the line profiles one could speculate that the
feature was present in some of the other spectra, but at different
energies. For example, P1 gives \delchi=6.0 for a feature at 5.8 keV,
significant at 95~per cent confidence. At this stage it is unclear
whether the feature is variable in strength, energy or both.  A
possible alternative explanation for the shape of the P4/P5 line is
that we are seeing two separate components, a core around 6.4 keV and
a strongly-redshifted and broad component at $\sim 5.5$~keV. The
latter is reminiscent of the profile occasionally seen in MCG-6-30-15
(I96).  The absorption interpretation seems preferable, however, as an
examination of, e.g., the P2 and P7 profiles is more suggestive of a
single line.

\section{DISCUSSION}

We have presented a broad, iron-line profile of NGC 3516 with
unprecedented signal-to-noise ratio. Disk models give an excellent fit
to the data, and indicate a strong concentration of the emission in
the close to the central black hole. We cannot distinguish between
rotating and non-rotating black hole models using these data alone,
although the disk-line parameters seem more plausible for a Kerr
geometry. There is also some evidence in the mean profile for a
narrower component which arises from somewhere other than the inner
accretion disk, and/or an absorption feature at around 5.9 keV. There
is also tentative evidence for the Fe K$\beta$ and/or Ni K$\alpha$
lines in the integrated spectrum. Consideration of the line
variability suggests that the profile is variable, meaning that the
fits to the integrated spectrum offer only a partial picture of the
central regions. The profile changes are not easily interpretable, but
energy-resolved light curves indicate that the core of the line
responds to the continuum, but that the extended wings do not. These
fluxes are also variable, however, and are correlated with each
other. The clearest interpretation of this is that the red and blue
wings form a single, broad component with a common origin in the inner
disk, but that a large fraction of the core comes from somewhat larger
radii.

The profile changes may be considered surprising given the
relatively weak continuum variability, and the inference that the bulk
of the flux in the line probably arises from within $100~R_{\rm g}$
(Tanaka et al. 1995; N97b). The light-travel time is only 5000
$R_{100} M_{7}$s (where $R_{100}$ is the radius in units of
$100~R_{\rm g}$ and $M_7$ the mass in units of $10^{7}$~\Msun, which
we consider a reasonable estimate; see, e.g., Edelson \& Nandra
1999). The line should therefore respond fully to any continuum
changes in the integrations presented in Fig.~\ref{fig:var_prof} and
they should simply exhibit the mean profile. If due to reverberation,
the profile variations would imply that we have seriously
underestimated either the size of the emission regions (e.g. Hua,
Kazanas \& Cui 1999) and/or the black hole mass.  Also, the variation
of the ``blue wing'' is in excess of a factor $2$, which is more than
the continuum (Fig.~\ref{fig:lc}).  Reverberation cannot produce such
an over-response without changes in ionization state, for which we
have no evidence. More likely the profile variations are dominated not
by reverberation, but some other process which acts in a different
manner or on a longer time scale.  For example the variation in the
blue wing could be due to an enhancement in the illuminating flux of
the inner disk portion moving towards us, due to a local flare
(c.f. I96, Iwasawa et al. 1999).  Strong gravitational effects in
transverse or receding portions at the same radius would account for
the correlation between the red wing and blue core. A more
``standard'' reverberation picture may be relevant for the core, which
comes from further out where the X-ray source appears point-like.  The
apparent response of the core to the continuum on time scales of $\sim
50$~ks still places that flux within $\sim 1000~R_{\rm g}$,
however.

Another physical process which can produce variations in the line
profile which are not necessarily related to flux and which can occur
on a longer time scale is absorption. Complete absorption
(i.e. occultation) has already been suggested as a possible mechanism
for profile variability (Weaver \& Yaqoob 1998).  We interpret our
absorption feature at $\sim 5.9$ keV, which is evident in the mean
spectrum but particularly strong in the P4 and P5 data, as being due
to resonance scattering by iron. Resonance absorption features are
expected, but unless there is a large velocity gradient in the
absorbing material they would typically not be observable in the
\asca\ spectra as they would be very narrow and therefore weak.  As
the feature is redshifted with respect to the rest energy of iron
K$\alpha$ emission (6.4-6.9 keV), we speculate that the material is
infalling and/or suffering gravitational redshift close to the central
hole. Evidence for infalling material in AGN is relatively scarce and,
if confirmed, such resonance absorption features could provide rare,
hard evidence for material actually accreting onto the black
hole. Differential (i.e. tidal) changes in the gravitational field
would naturally provide a large, apparent velocity gradient in the
material, broadening and strengthening the resonance features.
Further support for an origin close to the central hole is provided by
the variability of the feature, which will be particularly interesting
to follow up with the high-resolution XRS calorimeter aboard ASTRO-E.

\section*{ACKNOWLEDGMENTS}

We thank the \asca\ team for their operation of the satellite, and the
\asca\ GOF at NASA/GSFC for helpful discussions. We acknowledge the
support of the Universities Space Research Association (KN, IMG,
TY). KN is supported through NASA ADP grant NAG5-7067. This research
has made use of data obtained through the HEASARC on-line service,
provided by NASA/GSFC.


\clearpage

\newpage
\begin{references}

\reference{en99} Edelson, R.A., Nandra, K.,
	1999, ApJ, 514, 682 
\reference{fa95} Fabian, A.C., Nandra, K., Reynolds, C.S., Brandt, W.N., 
		 Otani, C., Tanaka, Y., Inoue, H., Iwasawa, K., 
 		   1995, MNRAS, 277, L11
\reference{fa89} Fabian, A.C., Rees, M.J., Stella, L., White, N.E.,
	1989, MNRAS, 238, 729
\reference{gf91} George, I.M., Fabian, A.C.,
	1991, MNRAS, 249, 352
\reference{ge98} George, I.M., Turner, T.J., Netzer, H., Nandra, K.,
		Mushotzky, R.F., Yaqoob, T., 
	1998, ApJS, 114, 73
\reference{gh93} Ghisellini, G., Haardt, F., Matt, G.,
		   1993, MNRAS, 267, 743  
\reference{hu99} Hua, X.-M., Kazanas, D., Cui, W.,
	1999, ApJ, 512, 793
\reference{iw96} Iwasawa, K., \etal,
	1996, MNRAS, 282, 1038 (I96)
\reference{iw99} Iwasawa, K., Fabian, A.C., Young, A.J., 
	Inoue, H., Matsumoto, C., 
	1999, MNRAS, 306, L19
\reference{ki76} Kikoin, I.K., 
	1976, Tables of Physical quantities, Atomizdat, Moscow
\reference{kr96} Kriss, G., et al.,
	1996, ApJ, 467, 629
\reference{kr94} Krolik, J.H., Madau, P., Zycki, P.,
		   1993, \apj, 420, L57
\reference{la91} Laor, A.,
		   1991, \apj, 376, 90
\reference{my98} McKernan, B., Yaqoob, T.,
	1998, ApJ, 501, L29
\reference{n97a} Nandra, K., George, I.M., Mushotzky, R.F., Turner, T.J., 
		Yaqoob, T.,
	1997a, ApJ, 476, 70
\reference{n97b} Nandra, K., George, I.M., Mushotzky, R.F., Turner, T.J., 
                 Yaqoob, T.,
	1997b, ApJ, 477, 602 (N97b)
\reference{n97c} Nandra, K., Mushotzky, R.F., Yaqoob, T., George, I.M., 
		Turner, T.J., 
	1997c, MNRAS, 284, L7 (N97c)
\reference{re97} Reynolds, C.S.,
	1997, MNRAS, 286, 513
\reference{rb97} Reynolds, C.S., Begelman, M.C.,
	1997, ApJ, 487, 135
\reference{re99} Reynolds, C.S, Young, A.J., Begelman, M.C., Fabian, A.C.,
	1999, ApJ, 514, 164 
\reference{ta95} Tanaka, Y., \etal,
	1995, Nature, 375, 659
\reference{wy98} Weaver, K., Yaqoob, T.,
	1998, ApJ, 502, 139
\reference{yo98} Young, A.J., Ross, R.R., Fabian, A.C., 
	1998, MNRAS, 300, L11
\end{references}
\end{document}